\documentclass[%
 reprint,
 amsmath,amssymb,
 aps,
 prb,
]{revtex4-1}

\usepackage{graphicx}% Include figure files
\usepackage{dcolumn}% Align table columns on decimal point
\usepackage{bm}% bold math
\usepackage{color}

\begin{document}

%\preprint{YITP}

\title{First Principles Calculation of Helical Spin Order in Iron Perovskite SrFeO$_3$ and BaFeO$_3$}% Force line breaks with \\

\author{Zhi Li$^1$$^{,}$$^{3}$, Robert Laskowski$^2$, Toshiaki Iitaka$^{3}$, and Takami Tohyama$^1$}
% \email{lunar@yukawa.kyoto-u.ac.jp}
\affiliation{%
$^1$Yukawa Institute for Theoretical Physics, Kyoto University, Kyoto 606-8502, Japan\\
$^2$Institute of Materials Chemistry, Vienna University of Technology, Getreidemarkt 9/165-TC, A-1060 Vienna, Austria\\
$^3$Computational Astrophysics Laboratory, RIKEN, 2-1 Hirosawa,
Wako, Saitama 351-0198, Japan
}%

%\author{Robert Laskowski}
%\affiliation{
%$^2$Institute of Materials Chemistry, Vienna University of Technology, Getreidemarkt 9/165-TC, A-1060 Vienna, Austria
%}%
%\affiliation{
%\author{Takami Tohyama}
%\affiliation{%
%Yukawa Institute for Theoretical Physics, Kyoto University, Kyoto 606-8502, Japan
%}%


\date{\today}% It is always \today, today,

\begin{abstract}
Motivated by recent discovery of ferromagnetism in cubic perovskite BaFeO$_3$ under small magnetic field, we investigate spin order in BaFeO$_3$ and isostructual SrFeO$_3$ by the first principles calculation. The on-site Coulomb and exchange interactions are necessary for the helical spin order consistent with experiments. SrFeO$_3$ exhibits stable G-type helical order, while A- and G-type helical orders in BaFeO$_3$ are almost degenerate at short propagating vector with tiny energetic barrier with respect to ferromagnetic spin order, explaining ferromagnetism under small field. The results are consistent with model calculation where negative charge-transfer energy is explicitly taken into account.
\begin{description}
\item[PACS numbers]
 75.30.-m, 75.30.Et, 75.50.Bb, 75.40.Mg
\end{description}
\end{abstract}

 %PACS, the Physics and Astronomy
\maketitle

%\tableofcontents
\section{Introduction}

The coexistence of spin order and metallic conductivity in 3$d$ transition-metal oxides have attracted a lot of research attention since the discovery of high-temperature superconductivity in copper oxides. The metallic conduction is accessible by the introduction of carriers into either Mott-Hubbard or charge-transfer insulators in the Zaanen-Sawatzky-Allen phase diagram.~\cite{Zaanen85} Without carrier doping, metallic conduction is expected in the phase diagram when systems are located in the area with small or negative charge-transfer energy $\Delta$. Such a situation can be obtained for large atomic number and high valence of 3$d$ transition metal.

A typical example of the negative $\Delta$ compounds is cubic perovskite SrFeO$_3$, where formal valence of iron is Fe$^{4+}$ (3$d^4$) and the effective value of $\Delta$ is estimated to be $\Delta\sim-3$~eV.~\cite{Bocuet92} Metallic conductivity has been preserved even below an antiferromagnetic (AFM) transition temperature $T_\mathrm{N}\sim 134$~K.~\cite{MacChesney65,Takeda00,Hayashi01} Below $T_\mathrm{N}$, SrFeO$_3$ shows a G-type helical spin order whose propagation vector is parallel to the [111] direction. The observed vector is $\mathbf{q}=0.112(1,1,1)\times 2\pi/a_\mathrm{S}$ with lattice parameter $a_\mathrm{S}=3.85$\AA.~\cite{Takeda72} The magnetic moment is 3.1$\mu_\mathrm{B}$ per iron at 4.2~K.~\cite{Takeda81} Recently versatile helimagnetic phase diagram under magnetic field has been established and  unconventional Hall resistivity has been discussed in connection with topological Hall effect in noncoplanar spin texture with scalar spin chirality.~\cite{Ishiwata11}

In contrast to SrFeO$_3$, CaFeO$_3$ is insulating below 290~K,~\cite{Kawasaki98} where a charge disproportionation by the formation of Fe$^{3+}$ and Fe$^{5+}$ emerges.~\cite{Takano77} The difference may come from the compression of lattice due to small ironic radius of Ca$^{2+}$ compared to Sr$^{2+}$. It is thus interesting to examine oppositely the effect of lattice expansion by replacing Sr$^{2+}$ by Ba$^{2+}$.

Very recently cubic BaFeO$_3$ has been synthesized by low-temperature chemistry.~\cite{Hayashi11} The ground state seems to be AFM and there is no charge disproportionation and no structural distortion down to 8~K, similar to SrFeO$_3$. However, its helical order is different from SrFeO$_3$: A-type helical order with a propagation vector along the [100] direction. Here, $\mathbf{q}=(0.06,0,0) \times 2\pi/a_\mathrm{B}$ with lattice parameter $a_\mathrm{B}=3.97$\AA. By applying small magnetic field $\sim 0.3$~T, the AFM state changes to ferromagnetic (FM) one whose saturated moment is 3.5~$\mu_\mathrm{B}$/Fe. This is also different from SrFeO$_3$ where saturated ferromagnetism (3.5~$\mu_\mathrm{B}$/Fe) is achieved by applying more than 40~T.~\cite{Ishiwata11}

The helical spin order in iron perovskites has been examined theoretically based on a double exchange model explicitly including oxygen 2$p$ orbital.~\cite{Mostovoy05} By adding superexchange interaction $J_\mathrm{SE}$ between localized spins, the G-type helical order is stabilized. Though the reduction of $J_\mathrm{SE}$ changes the G type to the A type in small and negative $\Delta$ region, it is not clear whether the change really corresponds to the difference of real materials between SrFeO$_3$ and BaFeO$_3$. In order to include material information in theory, we need to perform the first principles calculation on magnetic structures in these compounds based on density functional theory (DFT).

In this paper, we perform non-collinear spin polarized DFT calculations in SrFeO$_3$ and BaFeO$_3$ as a function of the propagation vector $\mathbf{q}$ for helical spin order. We find that the DFT calculation within local spin density approximation (LSDA) requires on-site Coulomb interaction $U$ and exchange interaction $J$ to explain experimentally observed propagation vectors.
The G-type helical order in SrFeO$_3$ is nicely reproduced by the LSDA+$U$ with $U=3$~eV, $J=0.6$~eV.  In BaFeO$_3$, although the G-type order is lower in energy, the energy difference between A- and G-type orders gets extraordinary small around experimentally observed propagating vector in contrast to SrFeO$_3$.
The energy difference between finite $\mathbf{q}$ and $\mathbf{q}=0$ (FM) orders is larger in SrFeO$_3$ than in BaFeO$_3$, consistent with experimental observation that SrFeO$_3$ requires a larger magnetic field to achieve saturated FM state. The difference of the two compounds is attributed to the difference of lattice constant. This is confirmed by performing a model calculation that includes both negative $\Delta$ and the difference of the lattice constant through the distance dependence of hopping and superexchang parameters.
We also find that the density of states (DOS) just below the Fermi level is dominated by oxygen 2$p$ component, consistent with the view of negative $\Delta$.

This paper is organized as follows. In Sec. II, we introduce the computational method for helical spin order using DFT. The results by DFT calculation are shown in Sec. III. In Sec. IV, we introduce a double exchange model, and show results consistent with the DFT calculations. We summarize our results from the DFT and model calculations in Sec. V.

\section{Computational method}

In order to treat helical spin order, we must constrain the form of wave function. \cite{Sandraskii86,Laskowski04} The wave function $\Psi_\mathbf{k}(\mathbf{r})$ under the propagation vector $\mathbf{q}$ could be written as

\begin{equation}
\Psi_\mathbf{k}(\mathbf{r})=
\left(
\begin{array}{cc}
e^{i\mathbf{(k-q/2)}\cdot\mathbf{r}} & 0 \\
0 & e^{i\mathbf{(k+q/2)}\cdot\mathbf{r}}
\end{array}
\right)
\left(
\begin{array}{c}
\mu_\mathbf{k}^\uparrow(\mathbf{r}) \\
\mu_\mathbf{k}^\downarrow(\mathbf{r})
\end{array}
\right)
,
\label{Psi}
\end{equation}
where $\mu_\mathbf{k}^\sigma(\mathbf{r})$ is translational invariant Bloch wave function with spin $\sigma$ before transformation. To determine the optimal propagation vector $\mathbf{q}$, we calculate the $\mathbf{q}$ dependent total energy $E(\mathbf{q})$ per unit cell by VASP code.~\cite{VASP} Non-collinear spin polarized calculations are performed within LSDA with and without $U$ and $J$.~\cite{Liechtenstein} The calculation is done in primitive cell with 10$\times$10$\times$10 $k$-points and energy resolution 0.01~meV per cell. Since the orbital moment is quenched completely by our calculation, the spin-orbital coupling effect is not under consideration. The cubic crystal structure (space group Pm-3m) is used. Crystal distortions may happen resulting from the spin lattice coupling, but this distortion is not detected by experiment down to 8 K. We thus consider that the distortion is very small and ignore it in our calculations.
\section{Results by DFT calculation}

\begin{figure}[t]
\includegraphics[width=7cm]{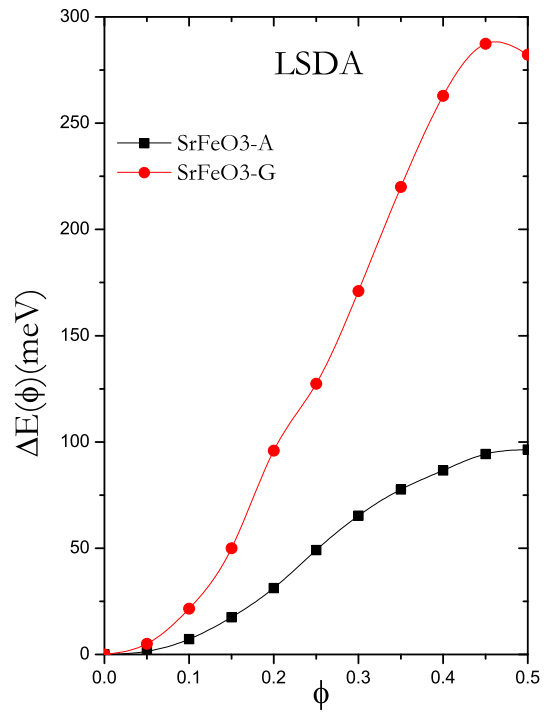}% Here is how to import EPS art
\caption{\label{fig1} (Color online) The $\phi$ dependence of total energy difference per unit cell, $\Delta E(\phi)\equiv E(\phi)-E(\phi=0)$, obtained by LSDA in SrFeO$_3$. For the definition of $\phi$, see text.}
\end{figure}

Figure~\ref{fig1} shows the total energy of SrFeO$_3$ as a function of the propagation vector by LSDA, where $\phi$ in the horizontal axis is defined by $\mathbf{q}=\phi(1,1,1)\times2\pi/a_\mathrm{S}$ for the G type and $\mathbf{q}=\phi(1,0,0)\times2\pi/a_\mathrm{S}$ for the A type with $a_\mathrm{S}=3.85$\AA.~\cite{Takeda72} The total energy is measured from the $\phi=0$ (FM) state: $\Delta E(\phi)\equiv E(\phi)-E(\phi=0)$. Within LSDA, the ground state is the FM state. If we take $U=3$~eV and $J=0.6$~eV, the minimum energy of SrFeO$_3$ shown in Fig.~2 is located at $\phi=0.11$, which agrees with the experimental findings. The magnetic field required to overcome the energetic barrier of $\Delta E(\phi=0.11)$=3.0~meV is approximately equal to 52~T, which is close to the experimental magnetic field ($\sim 42$~T) at which the FM moment is saturated.~\cite{Ishiwata11} The calculated moment at $\phi=0.11$ inside the muffin-tin spheres is 3.0$\mu_\mathrm{B}$/Fe, close to the experimental value of 3.1$\mu_\mathrm{B}$/Fe.~\cite{Takeda81}

\begin{figure}[t]
\includegraphics[width=7cm]{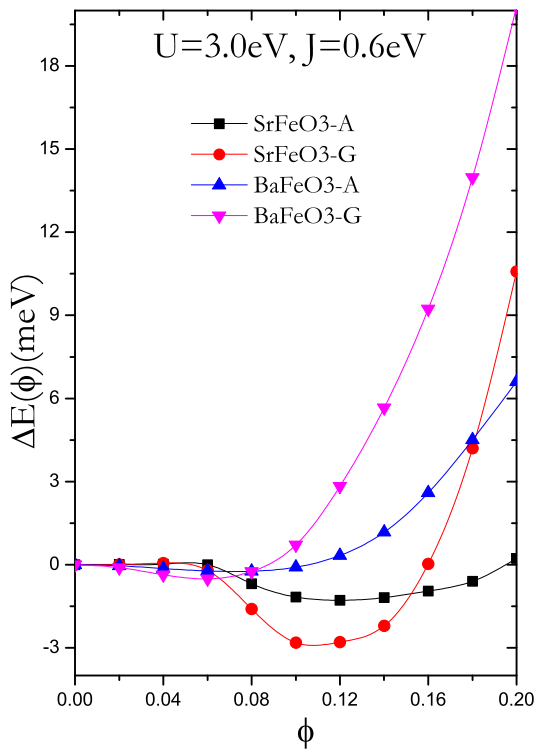}% Here is how to import EPS art
\caption{\label{fig2} (Color online) The $\phi$ dependence of total energy difference per unit cell, $\Delta E(\phi)\equiv E(\phi)-E(\phi=0)$, obtained by LSDA+$U$ in SrFeO$_3$ and BaFeO$_3$. For the definition of $\phi$, see text.}
\end{figure}

$\Delta E(\phi)$ for BaFeO$_3$ with lattice constant~\cite{Hayashi11} $a_\mathrm{B}=3.97${\AA}  is shown in Fig.~\ref{fig2}, where the same $U$ and $J$ are employed. We find that the total energy is almost independent of wave vector in both A- and G-type helical spin orders when $\phi<0.08$. This is consistent with the fact that the magnetic field to obtain saturated FM state is smaller~\cite{Hayashi11} ($\sim$1~T) than that in SrFeO$_3$.
The energy difference between A- and G-type is tiny, so the A-type helical spin order may be stabilized by, for example, introducing correlation effect that is beyond the present calculation. The small energy difference may affect physical properties at finite temperatures.

\begin{figure}[t]
\includegraphics[width=7cm]{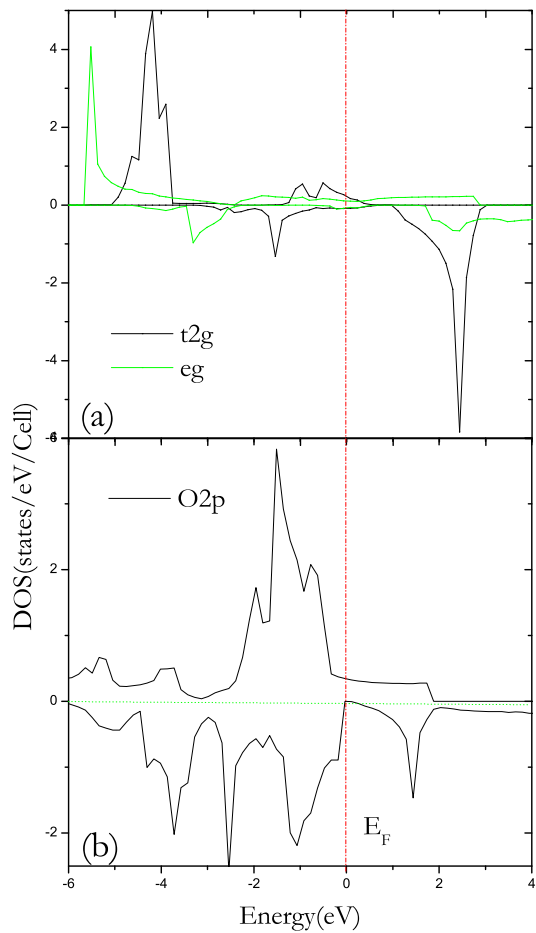}% Here is how to import EPS art
\caption{\label{fig3} (Color online) Density of states (DOS) of BaFeO$_3$ in the FM state calculated by LSDA+$U$ with $U=3.0$~eV and $J=0.6$~eV. The positive side of DOS denotes the up-spin DOS, while the negative side denotes the down-spin DOS. Dotted vertical line at zero energy represents the Fermi energy $E_\mathrm{F}$.
(a) DOS of Fe 3$d$ $t_{2g}$ and $e_g$ orbitals. (b) DOS of O 2$p$ orbitals.}
\end{figure}

In order to clarify the electronic structure of BaFeO$_3$, we show in Fig.~\ref{fig3} the DOS for the FM state ($\phi=0$), since this state is very close to the stable helical state.
Negative DOS represents down-spin component. The 2$p$ electrons of oxygen and 3$d$ of iron hybridize intensely in energy scale $-4\sim0$~eV, and spin-up component crosses the Fermi level, which indicates BaFeO$_3$ is half-metallic. The fact that DOS near the Fermi level is dominated by oxygen 2$p$ is consistent with the view of small or negative $\Delta$ in high valance 3$d$ transition-metal oxides. We also find that the $t_{2g}$ orbitals are almost occupied by up-spin electrons but empty for down spins. This indicates a localized nature of $t_{2g}$ spins.

What is the origin of different stable helical structures in SrFeO$_3$ and BaFeO$_3$? The most significant effect may come from lattice parameter from the results of our DFT calcination. Since ionic radius of Ba$^{2+}$ is larger than Sr$^{2+}$, BaFeO$_3$ can be considered as expanded SrFeO$_3$. We have confirmed that the energy difference between G type and A type in SrFeO$_3$ decreases with increasing the lattice parameter (not shown).
To get better understanding on the helical spin order and the effect of lattice in SrFeO$_3$ and BaFeO$_3$, we connect our DFT calculation with a double exchange model in Sec. IV.
\section{Results by model calculation}

Mostvoy~\cite{Mostovoy05} has shown that a double exchange model including explicitly oxygen 2$p$ orbitals bears the A-type helical spin order of localized $t_{2g}$ spins under small or negative $\Delta$. By adding superexchange interaction $J_\mathrm{SE}$ between the localized $t_{2g}$ spins, the G-type helical order emerges. The Hubbard $U$ in our first principle calculations is crucial for stabilizing the $t_{2g}$ localized spins so as to be consistent with the double exchange model. This is realized in Fig.~\ref{fig3} where all of up-spin electrons of the $t_{2g}$ are deep in energy. In addition, we expect superexchange process by making use of $U$, though the precise estimate of the process is difficult.
We note that small or negative $\Delta$ is necessary for the double exchange model to produce both the G- and A-type orders. In our LSDA+$U$ calculations, the presence of O2$p$ DOS just below the Fermi level (see Fig.~\ref{fig3}(b)) will be an indirect evidence of small or negative $\Delta$.

In order to make clear the correspondence between the double exchange model with $J_\mathrm{SE}$ and our LSAD+$U$, we have to show, at least, that the double exchange model can explain the difference in SrFeO$_3$ and BaFeO$_3$. For this purpose, we examine the model following Mostvoy's procedure.~\cite{Mostovoy05} By assuming that the localized $t_{2g}$ spin at site $i$ is in the $y$-$z$ plane, i.e., $\mathbf{S}_i=S(\hat{\mathbf{y}}\sin\mathbf{q}\cdot\mathbf{x}_i + \hat{\mathbf{z}}\cos\mathbf{q}\cdot\mathbf{x}_i)$ with unit vectors $\hat{\mathbf{y}}$ and $\hat{\mathbf{z}}$ along the $y$ and $z$ directions, respectively, the effective Hamiltonian of the model may be written as~\cite{Mostovoy05}

\begin{eqnarray}
H &=& \sum_{\mathbf{k},\alpha,\delta} t_{\alpha,\delta}\left( d_{\mathbf
{k}\alpha}^\dagger P_{\mathbf{k}\delta\downarrow} + \mathrm{h.c.} \right)
 + \Delta \sum_{\mathbf{k},\delta,\sigma} P_{\mathbf{k}\delta\sigma}^\dagger P_{\mathbf{k}\delta\sigma} \nonumber \\
&& + J_\mathrm{SE} \sum_{\left<i,j \right>}\mathbf{S}_i\cdot\mathbf{S}_j ,
\label{H_DE+SE}
\end{eqnarray}
with
\begin{equation}
P_{\mathbf{k}\delta\sigma}=2\left( \cos\frac{\mathbf{q}_\delta}{4} \cos\frac{\mathbf{k}_\delta}{2} p_{\mathbf{k}\delta\sigma} - \sin\frac{\mathbf{q}_\delta}{4} \sin\frac{\mathbf{k}_\delta}{2} p_{\mathbf{k}\delta
-\sigma} \right)
\label{P}
\end{equation}
where $\alpha$ runs two $e_g$ orbitals of $3z^2-r^2$ and $x^2-y^2$, $\delta$ runs three component of $x$, $y$, and $z$, $\left<i,j \right>$ runs the nearest-neighbor pairs, $d_{\mathbf{k}\alpha}$ represents the annihilation of $e_g$ electron, and $p_{\mathbf{k}\delta\sigma}$ represents the annihilation of 2$p_\delta$ electron with spin $\sigma$. The hopping matrix element $t_{\alpha,\delta}$ is given by $t_{3z^2-r^2,x}=t_{3z^2-r^2,y}=-(pd\sigma)/2$, $t_{3z^2-r^2,z}=(pd\sigma)$, $t_{x^2-y^2,x}=-t_{x^2-y^2,y}=\sqrt{3}(pd\sigma)/2$, and $t_{x^2-y^2,z}=0$, with the hopping parameter $(pd\sigma)$. We ignore oxygen-oxygen hopping for simplicity. We find from (\ref{P}) that finite $\mathbf{q}$ mixes both down and up spins for holes on oxygen.

\begin{figure}[t]
\includegraphics[width=7cm]{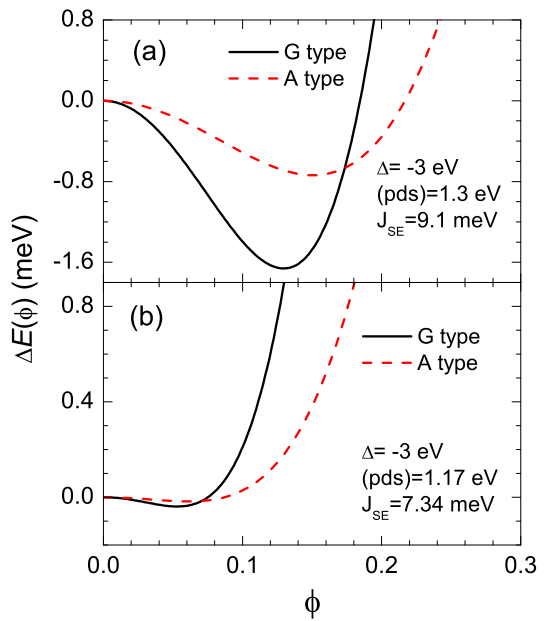}
\caption{\label{fig4} (Color online) The $\phi$ dependence of the total energy difference relative to the $\phi=0$, $\Delta E(\phi)$, of the model (\ref{H_DE+SE}). (a) SrFeO$_3$ and (b) BaFeO$_3$. Parameters denoted in each panel are taken in order for the minimum of the energy to be located at experimentally observed $\phi$ (see text). The G(A)-type helical spin order is represented by solid black (broken red) line. }
\end{figure}

We calculate the total energy of the Hamiltonian (\ref{H_DE+SE}) with one $e_g$ hole~\cite{Mostovoy05} and plot it as a function of $\phi$ for both the G- and A-type helical orders as shown in Fig.~\ref{fig4}. For SrFeO$_3$, we take $\Delta=-3$~eV and $(pd\sigma)=1.3$~eV according to photoemission analysis. \cite{Bocuet92} We determine a remaining unknown parameter $J_\mathrm{SE}$ in order for the energy minimum to be located at the experimentally observed propagation vector, as shown in Fig.~\ref{fig4}(a). $J_\mathrm{SE}$ is, then, estimated to be $J_\mathrm{SE}=9.1$~meV. Using the same parameter set, we find that the minimum energy of the A-type order is higher by 0.92~meV and the energy of the FM state ($\phi=0$) is also higher by 1.66~meV. The latter energy corresponds to the magnetic field of 28.7~T that is slightly smaller than the FM transition field $\sim$42~T for SrFeO$_3$.~\cite{Ishiwata11}

For BaFeO$_{3}$, we use a slightly smaller $(pd\sigma)=1.17$~eV taking the lattice expansion into account. By minimizing the energy at around experimentally observed $\phi$, we find $J_\mathrm{SE}=7.34$~meV, which is smaller than that for SrFeO$_3$ and is a reasonable value as a result of the expansion of the lattice constant compared with SrFeO$_3$. The energy relative to $\phi=0$ is very small with 0.017~meV as shown in Fig.~\ref{fig4}(b), which corresponds to 0.3~T that is comparable to the experimental value of 1~T. Such a small energy difference mainly comes from the decrease of $(pd\sigma)$ as compared with SrFeO$_3$. We note that the reduction of $J_\mathrm{SE}$ makes the energy difference between the G- and A-type orders considerably small, and the difference becomes almost zero for BaFeO$_3$.

We realize that the results of the model calculation are qualitatively consistent with the LSDA+$U$ results. Quantitative differences may be due to many assumptions for the construction of the double exchange model. Nevertheless, the agreement between the model calculation and LSDA+$U$ is satisfactory. This means that the LSDA+$U$ results contain the essential part of the model calculations, i.e., negative $\Delta$ and competition between the double exchange and superexchange processes.

\section{Summary}

We have examined helical spin order in cubic perovskite SrFeO$_3$ and BaFeO$_3$ by the first principles calculations based on DFT. We have found that $U$ and $J$ is necessary for explaining experimentally observed propagation vectors of spin. Including $U$ and $J$ to LSDA, we have obtained the G-type helical order for SrFeO$_3$ and almost degenerate A-type and G-type spin order for BaFeO$_3$. The energy difference between the minimum-energy state and FM state is larger in SrFeO$_3$ than in BaFeO$_3$, consistent with experimental observation that SrFeO$_3$ requires larger magnetic field to achieve saturated FM state. The difference of the two compounds is attributed to the difference of lattice constant. We have shown clear correspondence between the first principles calculation based on LSAD+$U$ and the double exchange model implemented by superexchange interaction, confirming the importance of the competitions between double exchange and superexchange interactions. The correspondence also implies that the characteristics of negative $\Delta$ is included in the first principles calculations. Correspondingly, the DOS of oxygen 2$p$ component in FM BaFeO$_3$ is just below the Fermi level, which is consistent with the view of negative $\Delta$.

\section*{Acknowledgment}
We would like to thank T. Morinari, K. Totsuka, N. Hayashi, H. Kageyama, M. Takano, and A. Fujimori for enlightening discussions. Zhi Li is grateful to the global COE program "Next Generation Physics, Spun from Universality and Emergence" and MEXT (No.20103005)for support. This work was also supported by Nanoscience Program of Next Generation Supercomputing Project, the Strategic Programs for Innovative Research (SPIRE), the Computational Materials Science Initiative (CMSI),the Grant-in-Aid for Scientific Research (Grants No. 22340097) from MEXT, and the Yukawa International Program for Quark-Hadron Sciences at YITP, Kyoto University. A part of numerical calculations was performed in the supercomputing facilities in YITP, Kyoto University, and RICC in RIKEN.

%\nocite{*}


\end{document}